\documentclass[
reprint,
superscriptaddress,
 amsmath,amssymb,
 aps,
pra,
]{revtex4-2}
\usepackage{graphicx}
\usepackage{dcolumn}
\usepackage{bm}
\usepackage{bbm}
\usepackage{physics}
\usepackage{hyperref}
\usepackage{amsmath}
\hypersetup{
     colorlinks   = true,
     linkcolor    = blue,
     citecolor    = blue,
     urlcolor     = blue
}

\begin{document}

\title{Quantum Period-Finding using One-Qubit Reduced Density Matrices}

\author{Marco Bernardi}
\email{bmarco@caltech.edu}
\affiliation{Department of Applied Physics and Materials Science, and Department of Physics,\\ California Institute of Technology, Pasadena, California 91125, USA}

\begin{abstract} 
The quantum period-finding (QPF) algorithm can compute the period of a function exponentially faster than the best-known classical algorithm. 
In standard QPF, the output state has a primary contribution from $r$ high-probability bit strings, where $r$ is the period. Measurement of this state, combined with continued fraction analysis, reveals the unknown period. 
Here, we explore a different approach to QPF, where the period is obtained from single-qubit quantities $-$ specifically, the set of one-qubit reduced density matrices (1-RDMs) $-$ rather than the output bit strings of the entire quantum circuit. 
Using state-vector simulations, we compute the 1-RDMs of the QPF circuit for a generic periodic function. Analysis of these 1-RDMs as a function of period reveals distinctive patterns, which allows us to obtain the unknown period from the 1-RDMs using a numerical root-finding approach. 
Our results show that the 1-RDMs $-$ a set of $\mathcal{O}(n)$ one-qubit marginals $-$ contain enough information to reconstruct the period, which is typically obtained by sampling the space of $\mathcal{O}(2^n)$ bit strings. Conceptually, this can be viewed as a ``compression'' of the information in the QPF algorithm, which enables period-finding from $n$ one-qubit marginals. 
Our results motivate the development of approximate simulations of reduced density matrices to design novel period-finding algorithms. 
\end{abstract}
\maketitle
%

%
%
\section{Introduction}
\vspace{-5pt}
Among quantum algorithms envisioned to achieve an advantage over classical computers, Shor's algorithm for integer factorization~\cite{shor1994,shor1997} provides a compelling example of how quantum computers could dramatically speed up calculations for useful tasks~\cite{Qadv, Boixo, Google, Zhong2020, Cirac2012,qsim-RMP,NISQ}. 
In Shor's algorithm, the factorization of a semiprime integer is turned into a period-finding problem~\cite{Chuang,Barnett}, where factoring an $n$-bit semiprime \mbox{integer} $S\!<\!N$, with $N\!=\!2^n$, requires finding the period of the modular exponential function, \mbox{$a^x \! \mod S$}, where $a$ and $S$ are relative primes. Finding this period requires $\mathcal{O}(2^n)$ resources on a classical computer: Since the period $r$ can take any value from 1 to $N-1$, a brute-force approach testing the possible periods would require on average $N/2 = 2^{n-1}$ attempts.
\\
\indent
In contrast, on a quantum computer, one can use the quantum period-finding (QPF) algorithm, which can find the period using resources that scale polynomially with the number of bits $n$ of the integer to be factored, achieving an exponential speed-up over its classical counterpart. 
The final step of the QPF algorithm measures the quantum state generated by the quantum Fourier transform. This state has a dominant component in a subspace spanned by $r$ high-probability bit strings, where $r$ is the period, as a result of interference and quantum correlations that are essential to achieve quantum speed up~\cite{Chuang}.
The QPF algorithm can be accurately modeled on a classical computer using state-vector simulations, but because of the $\mathcal{O}(2^n)$ size of the Hilbert space, such state-vector simulations perform similarly to the classical algorithm based on brute-force period testing, and are currently limited to $\sim$50 qubits with state-of-the-art classical computers~\cite{juqs,qhipster,quest,64qubit}. 
\\
\indent
Approximate methods for classical simulation of quantum circuits (QCs) provide an alternative to exact state-vector simulations. Examples include QC simulations based on tensor networks~\cite{Shi-TN, QFT, PRX}, density matrices~\cite{chen2021low,li2020density}, and neural-network quantum states~\cite{Carleo, gao-NN}, among others~\cite{Clifford,Vidal,qs-DMRG}. 
In particular, reduced density matrices (RDMs) can capture multi-qubit correlations and are smaller objects than state vectors: The combined size of the $s$-qubit RDM ($s$-RDM) matrices is $\mathcal{O}(n^s)$ for a QC with $n$ qubits. 
These RDMs can be computed exactly starting from the state vector, with exponential cost $\mathcal{O}(2^n)$, while approximate RDMs can be obtained at lower cost using approaches based on open system dynamics~\cite{lindblad1976,breuer2002, manzano2020}. 
Recent work has proposed efficient schemes for approximate simulations of 1-RDMs, obtaining accurate results for specific random QCs and oracle-based QCs~\cite{qcdft}. 
However, the structure of exact 1-RDMs (and higher $s$-RDMs) is not known for most quantum algorithms. This knowledge could shed light on quantum correlations and advance classical simulations of QCs. 
\\
\indent 
Here, we study the exact 1-RDMs of the QPF algorithm and discover a regular peak structure as a function of qubit and period. 
Knowledge of this pattern allows us to derive an approximate expression for the diagonal elements of the 1-RDMs, as a function of period for each qubit, valid for a generic periodic function. We show how to find the unknown period using a root-finding \mbox{algorithm} that compares the computed and approximate 1-RDMs. Similar to continued fractions in standard QPF, our method requires 2$n$ qubits to find an $n$-bit period with unit accuracy.
This approach \mbox{reconstructs} the period from $n$ one-qubit marginals of the QC distribution, and is conceptually different from standard QPF, where the period is obtained by sampling bit strings in a subspace of approximate size $r$, which scales as $2^n$ for a generic period.  
This work focuses on exact 1-RDMs, and thus finding the period still requires exponential resources.  
Yet, the approach developed here may inspire more efficient classical simulations of QPF and Shor's algorithm using approximate RDMs. We plan to investigate this topic in future work. 
\vspace{-10pt}
\section{Theory}
\subsection{Reduced density matrices}
\vspace{-10pt}
We study a periodic function $f(x) \!=\! f(x+r)$ defined for integers $x \in [0,2^n-1]$, where $n$ is the number of qubits, $N=2^n$ is the size of the domain, and the period $r$ can range from 1 to $N-1$.  
The 1-RDM for qubit $q$ is a $2\times2$ matrix obtained by tracing out $n-1$ qubits (all qubits $k$ except $q$) from the QC density matrix, $|\Psi\rangle \langle \Psi|$, where $|\Psi\rangle$ is the QC state-vector. The entire set of 1-RDMs, one per qubit, can be computed using 
\begin{align}
\label{eq:1rdm}
\begin{split}
\!\!\rho^{(q)}_{ij} &= \rm{Tr}_{\{k \} \ne q} (|\Psi\rangle \langle \Psi |) \\ & = \!\!\!\sum_{\{b_k\} \ne b_q}\!\! \langle b_{n-1} ...\, b_q \!=\! i\, ...\, b_0|\Psi\rangle \,
 \langle \Psi | b_{n-1}\, ...\, b_q \!=\! j\, ...\, b_0\rangle 
\end{split}
\end{align}
where $|b\rangle = |b_{n-1}\ldots b_1 b_0\rangle \in \{0,1\}^n$ are binary strings corresponding to integers $b=\sum_{l=0}^{n-1} b_l 2^{l}$ in $[0,2^n-1]$ (where $b_l = 0~\mathrm{or}~1$). The indices $i$ and $j$, equal to 0 or 1, label the elements of the 1-RDM matrices.
\\
\indent
For our problem involving periodic functions, we study the structure of the \mbox{1-RDMs} as a function of period $r$ for each qubit $q$ in the QC, focusing on $\rho^{(q)}(r)$ as the key quantities of interest. 
The 1-RDMs are Hermitian matrices with unit trace, so they can be written as 
\begin{equation}
\label{eq:as}
\rho^{(q)}(r)= \frac{\mathbbm{1}}{2} +  \bm{\sigma} \cdot \bm{a}^{(q)}(r)  = \begin{pmatrix}
0.5 + a_z & a_x -i a_y \\
a_x + i a_y & 0.5 - a_z 
\end{pmatrix},
\end{equation}
where $\bm{a}^{(q)}(r)=\{a_x, a_y, a_z\}$ are real-valued  coefficients for each qubit and period, here satisfying $2|\bm{a}^{(q)}(r)|=1$ (pure state), and $\bm{\sigma}=\{\sigma_x, \sigma_y, \sigma_z\}$ are Pauli matrices. 
The coefficients in the diagonal elements of the 1-RDMs, $a_z^{(q)}(r)$, are analyzed in detail in this work. 

\subsection{Quantum period-finding algorithm}
\begin{figure}[t]
\includegraphics[width=\linewidth]{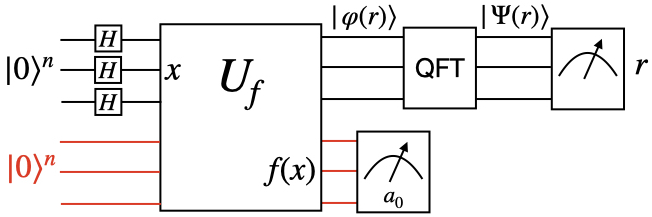}
\caption{Quantum circuit for QPF, shown for $n=3$ qubits in both registers. The oracle prepares the periodic function $f$ in the second register, shown at the bottom in red. Measuring the second register puts the first register in state $|\varphi(r)\rangle$, which  becomes $|\Psi(r)\rangle$ after the quantum Fourier transform (QFT) (see text for definitions). The final measurement, combined with continued fraction analysis, provides the period $r$.}
\label{fig:QC}
\end{figure}

\vspace{-10pt}
In the QPF circuit (see Fig.~\ref{fig:QC})~\cite{Chuang,Barnett}, the periodic function $f$ is prepared using two registers, each with $n$ qubits. After preparing the superposition state $H^{\otimes n} |0\rangle$ in the first register, where $H$ is the Hadamard gate, an oracle prepares the function $f$ in the second register. Following a measurement of the function in the second register, which returns $f(x)=a_0$, the state in the first register becomes~\cite{Barnett}:
\begin{equation}
\label{eq:beforeQFT}
|\varphi (r)\rangle = \frac{1}{\sqrt{\lceil N/r \rceil}} \sum_{m=0}^{\lfloor N/r \rfloor} |a_0 + mr\rangle, 
\end{equation}
where $\lfloor N/r \rfloor$ and $\lceil N/r \rceil$ are respectively the floor and ceiling of the ratio $N/r$. Applying the quantum Fourier transform on the first register generates the output state
\begin{equation}
\label{eq:afterQFT}
|\Psi (r)\rangle = \sum_{b=0}^{N-1}\,\left( \frac{1}{\sqrt{N\, \lceil N/r \rceil}} \sum_{m=0}^{\lfloor N/r \rfloor} e^{i 2\pi\, (a_0 + mr) \frac{b}{N}}\right) \, |b\rangle. 
\end{equation}
(Above, when $N/r$ is an integer, $\lfloor N/r\rfloor$ should be replaced by $N/r -1$.)
Measurement of this state in the computational basis, combined with continued-fraction analysis~\cite{Chuang,Barnett}, provides the unknown period $r$. Writing Eq.~(\ref{eq:afterQFT}) as $|\Psi(r)\rangle = \sum_b C_b(r) |b\rangle$, it is clear that the final measurement will return a bit string $b$ with probability $|C_b(r)|^2$, which is independent of the value $a_0$ measured in the second register~\cite{Barnett}. 
%
%
\subsection{1-RDMs for quantum period-finding}
\vspace{-10pt}
The diagonal elements of the 1-RDM play a central role in our treatment. Using Eq.~\eqref{eq:1rdm} on the QC output state $\lvert \Psi(r) \rangle$ defined in Eq.~\eqref{eq:afterQFT}, we obtain the diagonal elements as
\begin{align}
\begin{split}
\label{eq:rho00} 
\rho^{(q)}_{00} (r)&= \!\sum_{\{b_k\}\ne b_q} \!\! \lvert \langle b_{n-1} ...\, b_q \!=\! 0\, ...\, b_0|\Psi\rangle \lvert^2 \\
&= \sum_{b \in \beta_q} \frac{1}{N \lceil N/r \rceil} \left \vert \sum_{m=0}^{\lfloor N/r\rfloor} \exp\left(i 2\pi m\frac{rb}{N}\right) \right \vert^2 \\
&= \sum_{b \in \beta_q} |C_b (r)|^2,
\end{split}
\end{align}
where we define $\beta_q$ as the set of binary strings with the digit $b_q$ equal to zero and all other digits equal to 0 or 1. In the last line, the coefficient $|C_b(r)|^2$ is the probability of measuring a bit string $b$ as the output of the QC and is a familiar quantity in Shor's algorithm~\cite{Chuang,Barnett}. 
For a generic period $r$, $|C_b(r)|^2$ is maximal when $rb/N$ is close to an integer, and thus for \lq\lq special'' high-probability strings $b^*$ satisfying $b^* \approx j\,(N/r)$, with $j\!=\!0,1 ... r-1$. This becomes an exact equality when the period is a divisor of the domain size $N\!=\!2^n$, and thus for $r\!=\!2^k$ with $0\le k<n$. 
Note also that $\rho_{00}^{(q)}(r)$ is independent of the value $a_0$ measured in the second register. 
\\
\indent
In contrast, the coherences $\rho_{01}^{(q)}(r)$ depend on $a_0$ and are therefore less important to find patterns in the \mbox{1-RDMs.} This is seen by using the \mbox{1-RDM} definition in Eq.~(\ref{eq:1rdm}):
\begin{equation}
\label{eq:rho01}
\rho_{01}^{(q)}(r) \!=\! \!\!\!\sum_{\{b_k\} \ne b_q}\!\! \langle b_{n-1} ...\, b_q \!=\! 0\, ...\, b_0|\Psi\rangle \,
 \langle \Psi | b_{n-1}\, ...\, b_q \!=\! 1\, ...\, b_0\rangle 
\end{equation}
from which one finds that the coherences involve two terms where the coefficients $C_b$ are summed over different sets of binary strings: 
\begin{widetext} 
\vspace{-16pt}
\begin{align}
\rho^{(q)}_{01}(r) = \sum_{b \in \beta_q} \frac{1}{N \lceil N/r \rceil} 
\left ( \sum_{m=0}^{\lfloor N/r\rfloor} \exp\left(i 2\pi (a_0+mr)\frac{b}{N}\right) \right )\, \left ( \sum_{m=0}^{\lfloor N/r\rfloor} \exp\left(-i 2\pi (a_0+mr)\frac{\xi(b)}{N}\right) \right ).
\end{align}
\end{widetext}
In the first term in parentheses, the summation runs over $b \in \beta_q$, the set of strings with digit 0 for qubit $q$, while the second term contains the complementary strings $\xi(b)$, which are identical to the respective strings $b$ but with the digit at qubit $q$ set equal to 1. Because the two terms sum over different strings, the phase factor containing $a_0$ does not cancel, and $\rho_{01}^{(q)}(r)$ depends on $a_0$.  
Our numerical tests (not discussed here) confirm that $\rho_{01}^{(q)}(r)$ oscillates rapidly as a function of period without showing a clear pattern.\\ 
\indent 
For this reason, in this work we focus on the diagonal elements $\rho^{(q)}_{00}(r)$, whose physical meaning is clear from Eq.~(\ref{eq:rho00}): $\rho_{00}^{(q)}(r)$ is the probability of measuring zero for a single-qubit measurement on qubit $q$ in the output of the QPF circuit, and can be written as the sum of probabilities $|C_b(r)|^2$ of measuring a bit string $b\in \beta_q$. Therefore, $\rho^{(q)}_{00}(r)$ is a single-qubit probability, also known as a one-qubit marginal. This quantity has been studied in recent mean-field simulations of QCs~\cite{qcdft}.\\

\section{Results}
In Sections~\ref{sec:A} and ~\ref{sec:B}, we show calculations of 1-RDMs for the QPF circuit, focusing on the diagonal elements $\rho_{00}^{(q)}(r)$. Section~\ref{sec:A} discusses the simple case where the period $r$ is a divisor of the domain size $N\!=\!2^n$. This corresponds to the case where $N/r$ is an integer, so the period equals $r=2^k$ with $0\le k<n$, and the summations in Eq.~(\ref{eq:rho00}) can be evaluated analytically. 
\mbox{Section~\ref{sec:B}} focuses on the remaining periods that are not divisors of the domain size. In this case, $N/r$ is not an integer and the 1-RDMs need to be computed numerically. 
In Section~\ref{sec:C}, we analyze the calculated $\rho_{00}^{(q)}(r)$ to identify patterns as a function of period and qubit number.
This allows us to derive an approximate expression for $\rho_{00}^{(q)}(r)$, valid for a generic periodic function, and develop a numerical algorithm to obtain the period from the \mbox{1-RDMs.} 

\subsection{1-RDMs for quantum period-finding:\\ Case $r=2^k$}\label{sec:A}
\vspace{-10pt}
For periods $r=2^k$, we can derive analytical expressions for the 1-RDMs because $N/r$ is an integer. We first evaluate the sum over $m$ in Eq.~(\ref{eq:rho00}):
\begin{align}
\begin{split}
\rho^{(q)}_{00} (r=2^k) &= \sum_{b \in \beta_q} \frac{1}{N \cdot N/r} \left \vert \sum_{m=0}^{N/r -1} \left( e^{i 2\pi \frac{rb}{N}} \right)^m \right \vert^2 \\
& = \frac{1}{N \cdot N/r} \sum_{b \in \beta_q} \left( \frac{1-e^{i2\pi b}}{1- e^{i 2\pi \frac{rb}{N}}} \right)^2 \label{eq:r2k}
\end{split}
\end{align}
We now need to evaluate the sum over $b\in\beta_q$ for each qubit $q$ in the first register of the QPF circuit. In the second line of Eq.~\eqref{eq:r2k}, the numerator in parentheses is always zero because $b$ is an integer, and the only terms contributing to the sum are those where the denominator is also zero, which requires that $rb/N$ is an integer. When this condition is satisfied, using L'Hopital's rule, each such term contributes $(N/r)^2$ to the sum~\cite{Barnett}.\\ 
\indent
Let us count how many terms have an integer ratio $rb/N$. Using $r=2^k$ and $N=2^n$, where $n$ is the number of qubits and $k\in[0,n-1]$, we write the ratio as:
\begin{equation}
\label{eq:ratio}
\frac{rb}{N} = \sum_{l=0}^{n-1} b_l 2^{l+k-n}
\end{equation}
where we expressed $b$ in binary notation, $b=\sum_{l=0}^{n-1}b_l 2^l$. 
From Eq.~(\ref{eq:ratio}), the ratio 
$rb/N$ is an integer only when all exponents $l+k-n \geq 0$, and thus for bit strings with 
\begin{equation}
\label{eq:bl}
b_l = 0~~\text{for}~l < n-k.
\end{equation}
There can be up to $2^k=r$ such strings because the last $k$ digits, $b_l$ with $l\geq n-k$, can take any value.
\\
\indent
Depending on the qubit $q$ being considered, the number of strings with integer ratio $rb/N$, for $b \in \beta_q$, can be equal to this maximum allowed value of $2^k$, or smaller than that.  
To count such strings for qubit $q$, we examine the binary strings in the set $\beta_q$, which have digit $b_q=0$. 

If $q<n-k$, Eq.~(\ref{eq:bl}) is satisfied, and thus there are $r\!=\!2^k$ nonzero terms in the sum over $\beta_q$ in Eq.~(\ref{eq:r2k}), each contributing $(N/r)^2$. Therefore, we obtain
\begin{equation}
\rho^{(q)}_{00} (r\!=\!2^k) = \frac{r}{N^2} \cdot r \left(\frac{N}{r}\right)^2 = 1~~~\text{for}~q < n-k.
\end{equation}

If $q \geq n-k$, the set $\beta_q$ contains only half of the $2^k$ strings satisfying Eq.~(\ref{eq:bl}), because $b_q=0$ for $q\geq n-k$. This leads to a total of $2^{k-1}=r/2$ nonzero terms in the sum over $\beta_q$ in Eq.~(\ref{eq:r2k}), and thus we obtain:
\begin{equation}
\rho^{(q)}_{00} (r\!=\!2^k)= \frac{r}{N^2} \cdot \frac{r}{2} \left(\frac{N}{r}\right)^2 =\, 0.5~~~\text{for}~q \geq n-k.
\end{equation}
In the following, we express the 1-RDMs in terms of the coefficients $\bm{a}$ defined in Eq.~\eqref{eq:as}. In particular, $a_z$ is related to the diagonal elements of the 1-RDMs via $\rho_{00} = 0.5 + a_z$. 
To study $a_z^{(q)}(r)$ as a function of period $r$ and qubit $q$, it is useful to summarize the results above for $r\!=\!2^k$ as:
\begin{align}
\begin{split}
\label{eq:az2k}
a_z^{(q)} (r=2^k)&= 0.5 ~~~\text{for}~ k = 0, 1, ...\,, n\, –\, q\, –\, 1 \\
a_z^{(q)} (r=2^k)&= 0 ~~~~~\text{for}~k = n\, –\, q,\,n\, –\, q \, +1, ...\,, n-1.\\
\end{split}
\end{align}
This is a special case of a more general pattern discussed below.\vspace{20pt}\\ 

\begin{figure*}[t]
\centering
\includegraphics[width=1.0\textwidth]{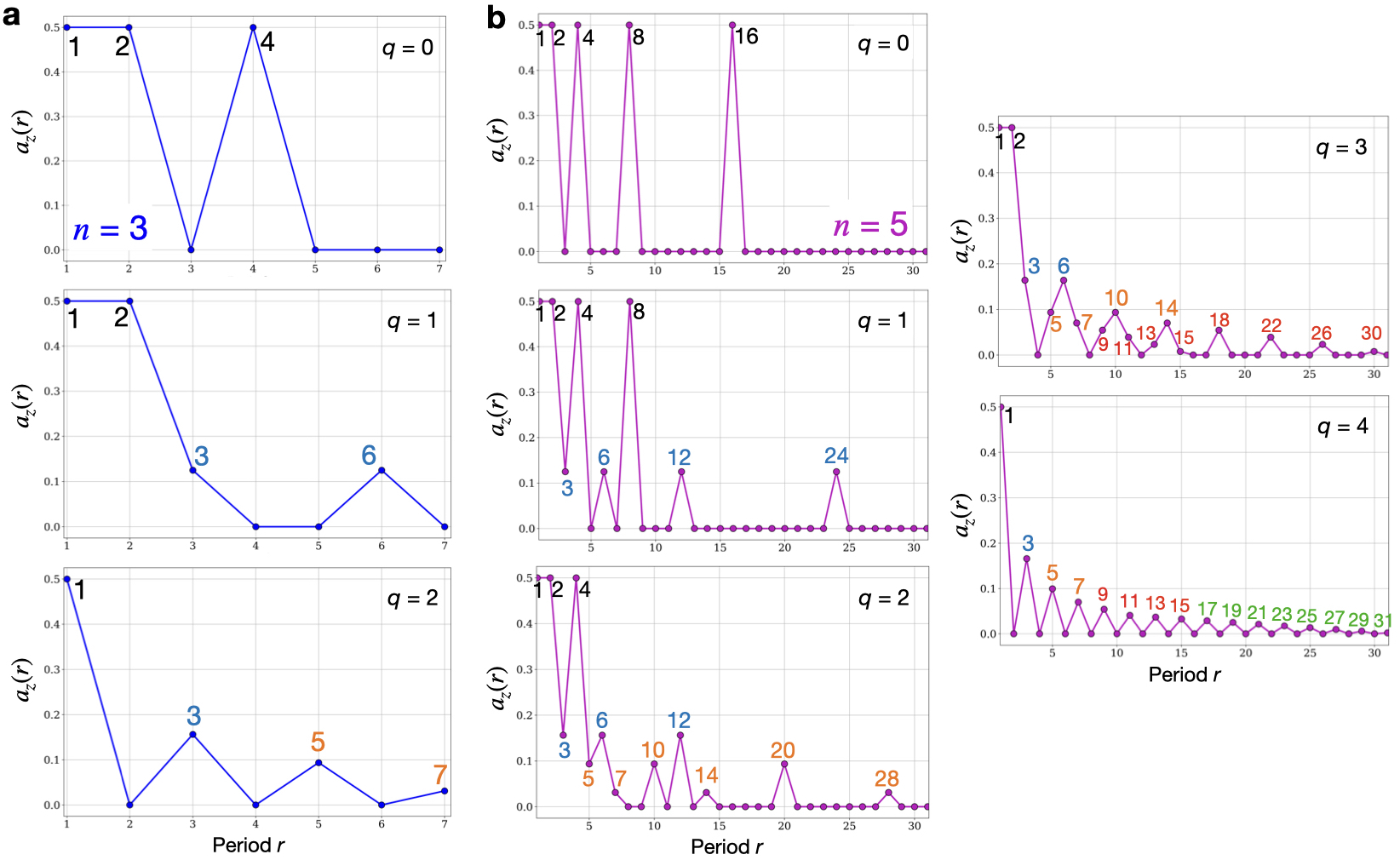}
\caption{Computed diagonal elements of the 1-RDMs for the QPF circuit, expressed as $a_z^{(q)}(r)  = \rho^{(q)}_{00}(r) - 0.5$, shown as a function of period $r$ for each qubit $q$. We show results for a) $n=3$ and b) $n=5$ qubits. The periods with $a^{(q)}_z(r)\!>\!0$ fit the general rule in Eq.~\eqref{eq:peaks}. For qubit $q=0$, a nonzero value of $a_z^{(q)}(r)$ occurs only for periods $r\!=\!2^k$. 
For each subsequent qubit ($q=1,2,...,n-1$), there are new periods with $a^{(q)}_z(r)\! >\! 0$, shown with colors not used in previous qubits, and removes some periods that had $a_z\!>\!0$ in preceding qubits. The last qubit, $q\!=\!n-1$, has $a_z(r) > 0$ only for odd periods.}
\label{fig2}
\end{figure*}

\subsection{1-RDMs for quantum period-finding:\\ Case $r\neq 2^k$}\label{sec:B}
\vspace{-10pt}
When $r\neq 2^k$, the period is not a divisor of the domain size $N\!=\!2^n$. 
Therefore, $rb/N$ is not an integer, so the sums in Eq.~\eqref{eq:rho00} cannot be evaluated analytically and the 1-RDMs need to be computed numerically. 
We compute the exact 1-RDMs using state-vector simulations of the QPF circuit followed by taking partial traces. 
The calculations loop over all possible periods and provide $\rho_{00}^{(q)}(r)$ for all qubits. 
In our analysis, we are only interested in the 1-RDMs for the $n$ qubits in the first register of the QPF circuit (see Fig.~\ref{fig:QC}), the one where the QFT is applied, while the 1-RDMs for the second register are ignored. 
The code and data sets for all simulations and 1-RDM calculations are provided in the Supplemental Material (SM)~\cite{SM}.
\\
\indent
To analyze the results, we focus on the physical meaning of $\rho_{00}^{(q)}(r)$. Since $\Tr(\rho) \!=\! \rho_{00} + \rho_{11} \!=\! 1$, when there is no constructive or destructive interference in the sum over strings in $\beta_q$ in Eq.~\eqref{eq:rho00}, one expects an average value of $\rho^{(q)}_{00}(r) = 0.5$, and thus $a^{(q)}_z(r) \!=\! 0$, because $\beta_q$ includes only half of the strings. Therefore, we can regard the value $a_z\!=\! 0$ as the baseline, and any value $a^{(q)}_z(r)\!>\!0$ as the result of constructive interference. 

Analysis of our computed 1-RDMs provides a pattern for the nonzero values of $a_z^{(q)}(r)$, which includes the analytic results in Eq.~(\ref{eq:az2k}) for $r\!=\!2^k$ as a special case, and generalizes them to all periods. 
For $r=2^k$, Eq.~(\ref{eq:az2k}) shows that $a_z^{(q)}(r)$ exhibits a series of peaks for periods equal to certain powers of 2. In particular, for a given qubit $q$, the value $a_z^{(q)}=0.5$ occurs for periods $r=2^k$ with $k = 0, ..., n-1-q$, while $a_z^{(q)}=0$ for periods equal to higher powers of 2. 
\\
\indent 
Using numerical results for the remaining periods, we are able to generalize this trend. Specifically, we find that $a^{(q)}_z(r)>0$ only for certain periods:
\begin{align}
\begin{split}
\label{eq:peaks}
a_z^{(q)} (r)&>0 ~~~\text{for}~ r= 2^k \!\cdot r', ~~\text{with odd}~r'< 2^{q+1}\\
&~~~~~~~~~~~~~~~~~~~~~~~~~~~~~\text{and}~~k=0,...,n\!-\!1\!-\!q.\\
a_z^{(q)}(r)&=0 ~~~\text{for all other periods}~r.
\end{split}
\end{align}\vspace{-10pt}
\\
\indent
To illustrate this trend, in Fig.~\ref{fig2}(a) we plot the results for $n\!=\!3$ qubits.
For qubit $q\!=\!0$, we find $a_z^{(q)}(r)=0.5$ for $r=\{1,2,4\}$, in agreement with Eq.~(\ref{eq:peaks}) for $r'=1$. 
Qubit $q\!=\!1$ introduces new periods with $a_z^{(q)}(r)>0$, including $r'=3$ and that period multiplied by powers of two, $2^k \!\cdot r'$, with $k=0,...,n-1-q$, consistent with Eq.~(\ref{eq:peaks}). Similarly, for qubit $q=2$, we find additional periods with nonzero $a_z$ values, the periods $r'=\{5,7\}$. Note that for the last qubit, $q=n-1$, the nonzero values of $a_z$ occur only for odd-valued periods, $r\!=\! 2^0 r'$, with odd $r'\!<\!2^n$, in agreement with the general trend given in Eq.~\eqref{eq:peaks}.
\\
\indent 
In Fig.~\ref{fig2}(b), we show the same pattern for the more complex case with $n=5$ qubits. Following our discussion above, a new set of periods with $a_z > 0$ will appear for each of the qubits $q=0,1,2$. These periods can be written as $2^k \cdot r'$, with $r'=1$ for qubit 0, $r'=3$ for qubit 1, and $r'=\{5,7 \}$ for qubit 2. 
In addition, there are new sets of periods with $a_z > 0$ for qubits $q=3,4$. These periods can also be written as $2^k \cdot r'$, with $r'=\{9,11,13,15\}$ for qubit 3, and all odd periods in the second half of the domain, $r'=\{ 17,19,...,31\}$ for qubit 4, which is the last qubit. This leaves only odd periods with $a_z>0$ for the last qubit, similar to the case with $n=3$. 
\\  
\indent
Based on these results, the trend, in terms of periods with $a_z > 0$, is that starting at $q=0$ each following qubit will:
i) add periods $2^k\!\cdot r’$, for $2^q \!<\! r’\!< 2^{q+1}$ ($2^{q–1}$ new odd periods $r’$, and these periods multiplied by powers of two); and ii) remove some periods with $a_z>0$ in previous qubits: specifically, remove periods $2^{n-q}\cdot r’$, with $r’ \!<\! 2^q$. We have verified that these trends are general and hold for larger circuit sizes up to $n > 10$. Although these results cannot be derived analytically since Eq.~(\ref{eq:rho00}) needs numerical calculations for generic periods, we can rationalize their origin, as we explain in the following.
\subsection{Finding the period from the 1-RDMs}\label{sec:C}
\vspace{-10pt} 
To find the unknown period from the 1-RDMs, we are given a set of $n$ values $a_z^{(q)}$, one for each qubit $q$. 
These values can be obtained from single-qubit measurements or classical simulations of the QPF circuit in Fig.~\ref{fig:QC}. The key question is whether the diagonal elements of the 1-RDMs $-$ the $n$ values $a_z^{(q)}$, which are one-qubit marginals of the QC distribution $-$ contain enough information to reconstruct the period. 
In the following, we show that these quantities are a unique fingerprint of the period. We then show how to obtain the period from the $n$ values $a_z^{(q)}$ (1-RDM diagonal elements) using a numerical root-finding approach. 
\begin{figure*}[t]
\centering
\includegraphics[width=1.0\textwidth]{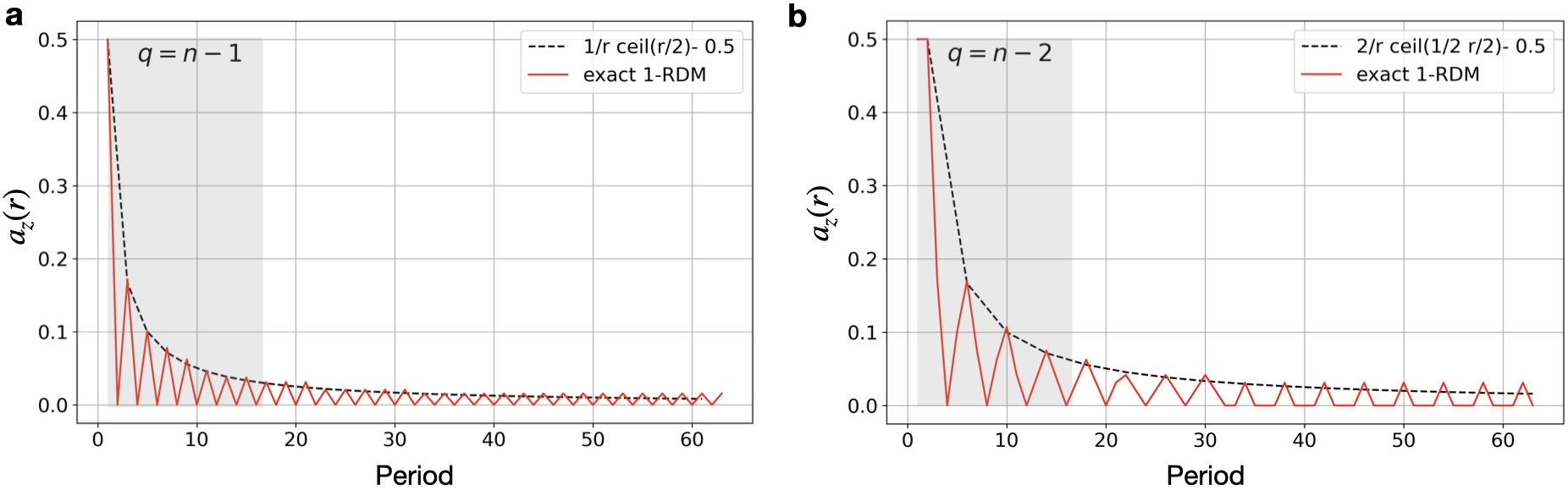}
\caption{Comparison of approximate versus exact $a^{(q)}_z(r)$ for $n\!=\!6$ qubits, shown for a) the last qubit, $q=n-1$, and b) the penultimate qubit, $q=n-2$. These cases correspond to $q'\!=\!0$ and $q'\!=\!1$, respectively. The first quarter of the domain (periods smaller than $N/4$), where the approximation is more accurate, is shown with a shaded area in both plots.}
\label{fig3}
\end{figure*}
\subsubsection{\bf Numerical approach for quantum period-finding from the 1-RDMs}
\vspace{-10pt}
Starting from the patterns discussed above, our strategy is to first derive an accurate approximation for the 1-RDM diagonal elements, $a_z^{(q)}(r)$, as a function of the period, and then compare the calculated $a_z^{(q)}$ with the approximate values to find the period. Let us denote the approximate 1-RDM diagonal elements as $\tilde{a}^{(q)}_z(r)$. 
To find the unknown period $r$ from the values $a^{(q)}_z$, 
we propose to solve for $r$ in the equation:
\begin{equation}
\label{eq:qpf}
a^{(q)}_z - \tilde{a}^{(q)}_z (r) = 0.
\end{equation}  
A possible approach would be to find the period that minimizes the distance between the calculated and approximate $a^{(q)}_z$. However, since the number of possible periods spans the entire domain, this period-finding approach would have an exponential computational cost of order $\mathcal{O}(2^n)$. 
To avoid this complexity, after deriving approximate $\tilde{a}^{(q)}_z (r)$, we treat Eq.~\eqref{eq:qpf} as a root-finding problem and use the secant method~\cite{Heath2018scientific} to solve numerically for the period $r$. The numerical approach is described in detail in the following. 
\\
\indent
The key to our method is the derivation of approximate 1-RDMs, $\tilde{a}^{(q)}_z(r)$. 
From the definition in Eq.~(\ref{eq:rho00}), $\rho^{(q)}_{00}(r) = \sum_{b\in\beta_q} |C_b(r)|^2$, we know that $\rho^{(q)}_{00}(r)$ is a sum of contributions from strings $b\!\in\!\beta_q$ that are approximately equal to an integer multiple of $N/r$. 
For the case $r=2^k$, this becomes an exact equality and $a^{(q)}_z(r)$ can be computed analytically, as discussed above. For a generic \mbox{period} $r$, we approximate $\rho^{(q)}_{00} (r)$ by counting the number of integer multiples of $N/r$ that fall within the range of values covered by the strings in $\beta_q$:
\begin{equation}
\label{eq:rho_approx}
\rho^{(q)}_{00} (r) \approx  \frac{1}{r} \sum_{j=0}^{r-1} \delta_{jN/r,\,b\in\mathcal{R}(\beta_q)} 
\end{equation}
where $\mathcal{R}(\beta_q)$ is the range of real values spanned by the strings in $\beta_q$, which consists of a set of intervals on the real number line (see below). 
This formula counts the number of values $b\in \mathcal{R}(\beta_q$) equal to $jN/r$, where $j\!=\!0,...,r-1$, and normalizes the result by $r$. 
When the number of matching values is greater than $r/2$, we expect $\rho_{00} > 1/2$, and thus $a_z > 0$. This shows that a value of $a_z\!>\!0$ is associated with constructive interference. 
\\
\indent
Counting the matching values in Eq.~\eqref{eq:rho_approx}, we obtain an approximate formula (see the derivation in the next section): For qubit $n-1-q'$, namely qubit $q'$ from the last, the approximate 1-RDM diagonal elements for the periods $r=2^{q'}\!\cdot\!r'$, with odd $r' < N/2^{q'}$, read
\begin{equation}
\label{eq:approx}
\tilde{\rho}^{(n-1-q')}_{00} (r) \approx \frac{2^{q'}}{r} \lceil{(r/2^{q'})/2}\rceil 
\end{equation}
where $\lceil x\rceil$ denotes the ceiling of $x$. Using $a_z = \rho_{00} - 0.5$, we obtain the approximate values $\tilde{a}^{(q)}(r)$ to be used in Eq.~(\ref{eq:qpf}):
\begin{equation}
\label{eq:approx2}
\tilde{a}^{(n-1-q')}_z (r) \approx \frac{2^{q'}}{r} \lceil{(r/2^{q'})/2}\rceil - 0.5\,. 
\end{equation}
\indent
Figure~\ref{fig3} compares this approximate formula with the exact ${a}^{(q)}_z(r)$ from state-vector simulations for the last two qubits in the first register of the QPF circuit. 
For qubit $n-1-q'$, the approximate formula can accurately predict the local peaks of $a^{(q)}_z (r)$, which occur for periods $r=2^{q'}\cdot r'$, with odd $r' < N/2^{q'}$. 
This is particularly true for periods in the first quarter of the domain, $r<N/4$, which allows us to find any period in a given domain by adding a few extra qubits to push the period of interest well into the first quarter. The accuracy of $\tilde{a}^{(q)}_z (r)$ in that region is the key to finding the period by solving Eq.~(\ref{eq:qpf}) numerically. 
\\
\indent
We describe the algorithm to find the period from the 1-RDM diagonal elements. The corresponding code is provided in the notebook included in the SM~\cite{SM}.  
\begin{enumerate}
\item 
\vspace{-3pt}
We first generate a list of periods compatible with the values of $a^{(q)}_z$. 
Starting at qubit 0, we find the first qubit where $a^{(q)}_z > 0$. If this occurs at qubit $l$, then the unknown period has the form $2^k\!\cdot r'$, with odd $r'$ such that $r' \in [2^l, 2^{l+1})$, for some value of $k$.   
\item 
\vspace{-2pt}
We then find the value of $k$ to restrict the list of possible periods. Starting at the last qubit, we find the first qubit from the last where $a^{(q)}_z > 0$. If this occurs for qubit $q'$ from the last, defined as qubit $q^*=n-1-q'$, then we can restrict the list of possible periods to $2^{q'}\!\cdot r'$, for odd $r' \in [2^l, 2^{l+1})$. 
\item 
\vspace{-2pt}
If the period is in the sets $2^k$ or $2^k \cdot 3$, then Steps 1$-$2 are sufficient to find the period. For other cases, we need to refine our search within the list of possible periods generated in the previous step. 
\item
\vspace{-2pt}
For this refinement, we examine $a^{(q)}_z$ for qubit $q^*$ defined above. Using Eqs.~(\ref{eq:qpf}) and (\ref{eq:approx2}), we solve for $r$ in $a^{(q*)}_z - \tilde{a}^{(q*)}_z(r) = 0$ using the secant method~\cite{Heath2018scientific} over the set of possible periods found in Step 2, starting at the midpoint of that range of periods. 
\item
\vspace{-2pt}
To validate the result, we verify that the period found with the secant method is in the set of possible periods from Step 2. As the procedure is more accurate when $r<N/4$, we then re-run the period-finding algorithm by adding more qubits, until the computed period converges. 
\end{enumerate}

\begin{figure}[t]
\centering
\includegraphics[width=1.0\linewidth]{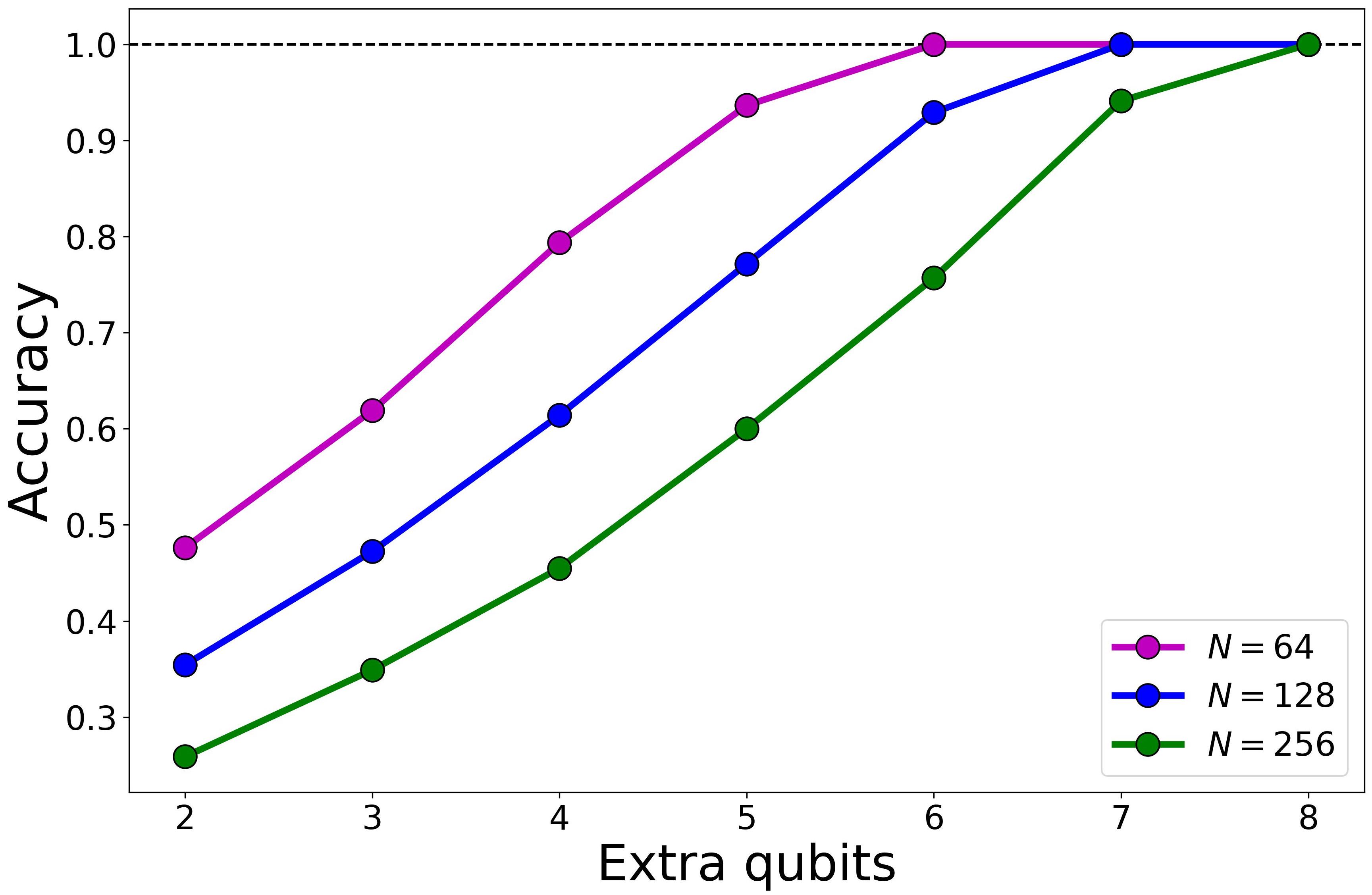}
\caption{Accuracy for quantum period-finding from 1-RDMs using our numerical approach. Results are shown for $N\!=\!64,~128,~256$, which corresponds to $n\!=\!6,~7,~8$ qubits respectively, by plotting the accuracy as a function of the number of extra qubits added to improve the root-finding procedure.}
\label{fig4}
\end{figure}

\indent
We use this approach to find periods in domains of integers with $n=6,~7$ and $8$ bits, respectively (periods up to \mbox{$N\!=\!256$}). Although only $n$ qubits are strictly necessary to find periods up to $2^n$, the use of additional qubits pushes the periods of interest to the early part of the domain, where our approximation for $a^{(q)}_z(r)$ is more accurate (see Fig.~\ref{fig3}). In Fig.~\ref{fig4}, we plot the accuracy of our method, defined as the fraction of periods in the domain predicted correctly, as a function of extra qubits. 
For $n$-bit integers, we find that using $n$ extra qubits, for a total of $2n$ qubits, allows us to predict the period with unit accuracy for all the domain sizes studied here. The requirement of $2n$ qubits to find the period with unit accuracy from the 1-RDMs is equivalent to the standard QPF algorithm based on bit strings $-$ for example, in Shor's algorithm $-$ which requires $2n$ qubits to find the period reliably using continued fractions~\cite{Chuang}. 

\subsubsection{\bf Derivation of 1-RDM approximate formula}
\begin{figure}[b]
\centering
\includegraphics[width=0.8\linewidth]{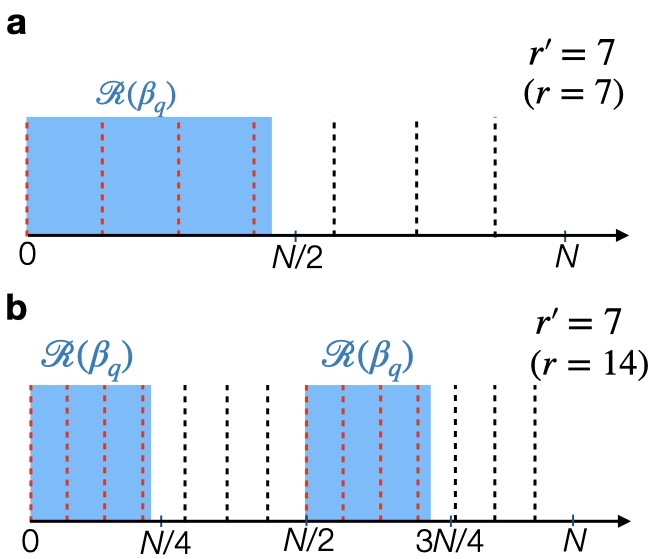}
\caption{Relation between the range $\mathcal{R}(\beta_q)$ and the values $jN/r$ for a) the last qubit $q\!=\!n-1$, for $r\!=\!7$, and b) the next to last qubit, $q\!=\!n-2$, for $r\!=\!2r'$, with $r'\!=\!7$. The range of real values spanned by the strings in $\beta_q$, $\mathcal{R}(\beta_q)$, is a set of continuous intervals, shown using blue color. The vertical dashed lines represent $jN/r$, for $j=0,1,...,r-1$, with the values overlapping with $\mathcal{R}(\beta_q)$ highlighted in red.}
\label{fig5}
\end{figure}

\vspace{-10pt}
According to Eq.~(\ref{eq:rho_approx}), $\rho^{(q)}_{00}(r) = a^{(q)}_z (r)+ 0.5$ counts the number of strings $b \in \beta_q$ that are approximately equal to $j N/r$, with $j=0,...,r-1$. To derive the approximate formula for $\tilde{\rho}_{00}^{(q)}(r)$ in Eq.~(\ref{eq:approx}), we focus on $\beta_q$, the set of integers corresponding to binary strings of length $n$ with digit $q$ equal to 0, and compare these integers with $j N/r$.  
\\  
\indent
We first analyze the last qubit, $q\!=\!n-1$, where for the exact 1-RDMs, $a^{(n-1)}_z(r) \!>\!0$ for odd periods and $a^{(n-1)}_z(r)\!=\!0$ for even periods in the domain [see Fig.~\ref{fig3}(a)]. 
We want to understand the origin of this trend to formulate approximate 1-RDMs. 
Writing the strings as $b=(b_{n-1}b_{n-2}...b_1 b_0)$, we have $\beta_q \!=\! \{(0\,b_{n-2}...b_1 b_0)\}$ for $q\!=\!n-1$, which corresponds to integers that span the first half of the domain, the interval from 0 to $N/2 -1$, with $N=2^n$. We define this interval as the range of real values spanned by $\beta_q$ and denote it by $\mathcal{R}(\beta_q)$.
\\
\indent
For odd periods, here denoted as $r'$, more than half of the values $j N/r'$ fall in the first half of the domain, which is spanned by $\mathcal{R}(\beta_q$). 
For example, consider the case $r'\!=\!7$, illustrated in Fig.~\ref{fig5}(a). The values $j N/r'$, with $j\!=\!0,...\, r'-1$, are $\{0,\, N/7,\, 2N/7,\, 3N/7,\, 4N/7,\, 5N/7,\, 6N/7 \}$, and thus 4 out of 7 values of $j N/r'$ are in the first half of the domain, where 4 equals the ceiling of $r'/2$.  
This shows that Eq.~(\ref{eq:approx}), for $q\!=\!n-1$ (and thus $q'\!=\!0$), holds in this \mbox{case:} 
\begin{equation}
\label{eq:last}
\tilde{\rho}^{(n-1)}_{00} (r') \approx \frac{1}{r'} \lceil{r'/2}\rceil = \frac{4}{7} > 0.5,~~~~r'\text{ odd}.
\end{equation}
The result $\tilde{\rho}^{(n-1)}_{00} (r') \!=\! 1/r' \lceil{r'/2}\rceil > 0.5$, and thus $a^{(n-1)}_{z}(r')\!>\!0$, applies to all odd periods for the last qubit. 
\\
\indent
In contrast, for even periods $r$, exactly half of the values $jN/r$ fall in the first half of the domain spanned by $\beta_q$. Consider the case $r=6$, with $j N/r = \{0,\, N/6,\, N/3,\, N/2,\, 2N/3,\, 5N/6 \}$. Only 3 of 6 values are in the range from 0 to $N/2-1$ spanned by $\beta_q$. 
Therefore,
\begin{equation}
\label{eq:lasteven}
\tilde{\rho}^{(n-1)}_{00} (r) \approx \frac{1}{r} \cdot \frac{r}{2} = 0.5,~~~~r\text{ even}.
\end{equation}
Together with Eq.~\eqref{eq:last}, this explains the pattern observed in the last qubit, where $a_z\!>\!0$ only for odd periods, while $a_z\!=\!0$ for even periods. 
\\
\indent
Next, we focus on the penultimate qubit, $q\!=\!n-2$. For that qubit, we do not examine odd periods as we can already find those using the last qubit. Rather, we focus on the new peaks $a^{(n-2)}_z (r) >0$ that are not present in the last qubit. These peaks occur for even periods, $r=2 r’$, with odd $r’ < N/2$, for which the approximation in Eq.~(\ref{eq:approx}) holds [see Fig.~\ref{fig3}(b)]. 
For these periods, we can write $j N/r \!=\! j (N/2)/r’$ with $r'$ odd, and repeat the reasoning above but now on half of the domain, with size $N/2$. 
For $q\!=\!n-2$, $\beta_q$ is the set of strings with digit 0 in the penultimate qubit. Therefore, $\beta_q = \{(b_{n-1}\,0\, b_{n-3}...b_0)\}$ spans the first and third quarters of the domain, as shown schematically in Fig.~\ref{fig5}(b). (For example, for 4 qubits, $N=2^n=16$, and $\beta_{q=n-2} = \{0,1,2,3,8,9,10,11\}$).
Consider the values $j (N/2)/r’$, with $j=0,1,...,2r'-1$. Since $r'$ is odd, more than half of these values will fall in the first and third quarters of the domain, which are spanned by $\mathcal{R}(\beta_q)$. 
\\
\indent
For example, consider the case $r\!=\!2r'\!=\!14$ ($r'\!=\!7$) in Fig.~\ref{fig5}(b). The values $j (N/2)/r’$ in the first quarter of the domain, $[0,N/4)$, are $(N/2)\cdot\{0,1/7,2/7,3/7 \}$, and those in the third quarter, $[N/2,3N/4)$, are \mbox{$(N/2)\cdot\{1,8/7,9/7,10/7\}$.} 
In both quarters, we have $\lceil r'/2 \rceil=4$ values of $jN/r$ inside $\mathcal{R}(\beta_q)$. Adding these contributions, from Eq.~(\ref{eq:rho_approx}) we get
\begin{equation}
\label{eq:penul}
\tilde{\rho}^{(n-2)}_{00} (r) \approx \frac{2}{r} \lceil (r/2)/2 \rceil > 0.5,~~~~r=2r'\text{, odd $r'<N/2.$}
\end{equation}
This is the same result as Eq.~\eqref{eq:last}, as can be seen by substituting $r=2r'$, and it agrees with the approximate formula for the 1-RDM diagonal elements, Eq.~(\ref{eq:approx}). 
\\
\indent
Finally, we generalize this formula to all qubits. In our algorithm, we use the qubit $q'$ from the last, $q\!=\!n-1-q'$, to find the periods $r\!=\!2^{q'}r'$, with $r'$ odd and smaller than $N/2^{q'}$. For that qubit, we have $jN/r = j(N/2^{q'})/r'$ and the integers in $\beta_q$ span $2^{q'}$ intervals of size $N/2^{q'+1}$, with the initial points of adjacent intervals spaced apart by $N/2^{q'}$. 
Because $r'$ is odd and smaller than $N/2^{q'}$, there are $\lceil r'/2 \rceil = \lceil{(r/2^{q'})/2}\rceil$ values of $jN/r$ in each of the intervals spanned by $\mathcal{R}(\beta_q)$, for a total of $2^{q'} \lceil{(r/2^{q'})/2}\rceil$ values of $jN/r$. Therefore, using Eq.~(\ref{eq:rho_approx}), we obtain
\begin{equation}
\tilde{\rho}^{(n-1-q')}_{00} (r) \approx \frac{2^{q'}}{r} \lceil{(r/2^{q'})/2}\rceil,
\end{equation}
which is exactly Eq.~(\ref{eq:approx}), the approximate formula for the 1-RDM diagonal elements we set out to derive. In our algorithm, we use this formula to approximate the peaks of $\rho^{(n-1-q')}_{00}(r)$ at periods $r\!=\!2^{q'}r'$, with odd $r' \!<\!N/2^{q'}$. This allows us to find arbitrary periods in the domain.\\

\section{Discussion}
Our treatment of QPF focuses on finding the period of a function using only single-qubit quantities, the diagonal elements of the 1-RDMs for all qubits in the first register of the QPF circuit. These one-qubit marginals, $\rho_{00}^{(q)}$, correspond to the probability of measuring 0 for qubit $q$ when conducting single-qubit measurements, and consist of a set of $n$ real numbers between 0 and 1. 
We have shown that these single-qubit quantities suffice to find the period of a generic periodic function. This approach is different from the standard QPF algorithm, which samples bit strings in the output of the QPF circuit and uses reduced fraction analysis to obtain the period. In both approaches, for a function defined in the domain of $n$-bit integers, one needs $\sim$2$n$ qubits in the first register to obtain the period reliably. 
\\
\indent
We briefly discuss two implications of these results. 
First, similarly to other quantum algorithms, the output state of the QPF circuit is concentrated around a set of high-probability bit strings. 
(In the case of QPF, the output is dominated by $r$ bit strings, where the period $r$ can take any value from 1 to $2^n-1$.) 
However, our results show that thinking in terms of the entire Hilbert space of bit strings is not necessary because the 1-RDM one-qubit marginals contain the same information about the period as the entire set of bit strings. 
Since the 1-RDM diagonal elements for $n$ qubits are only $n$ numbers, as opposed to $r \approx \mathcal{O}(2^n)$ bit strings, conceptually our approach can be viewed as a \lq\lq compression'' of the information in the QPF algorithm. 
\\
\indent 
Second, our approach works for generic periodic functions, including the modular exponential function used in Shor's quantum algorithm for integer factorization. Since calculating \textit{exact} 1-RDMs from state-vector simulations requires exponential resources that scale as $\mathcal{O}(2^n)$, there is no benefit in extracting the period from exact 1-RDMs as opposed to bit strings sampled from state-vector simulations. However, \textit{approximate} simulations able to obtain the 1-RDMs with high accuracy may become available in the future, for example using methods based on tensor networks or open quantum systems. 
Combined with the approach described in this work, hypothetical classical simulations $-$ not yet available to our knowledge $-$ capable of predicting accurate 1-RDM marginals in polynomial time for the QPF circuit may enable more efficient classical algorithms for period finding and integer \mbox{factorization.} 
\\
\indent
In summary, we show an approach to find the period of a function in the domain of $n$-bit integers, a quantity with formal exponential complexity of $\mathcal{O}(2^n)$, using only $n$ one-qubit marginals. 
This is achieved by finding patterns in the 1-RDMs, formulating reliable approximations, and developing a numerical approach to obtain the period from the 1-RDMs. 
Future work will focus on the development of accurate approximations for one- and multi-qubit RDMs for QPF and other quantum \mbox{algorithms.} 
\vspace{10pt}
\begin{acknowledgments}
M.B. gratefully acknowledges Vittorio Giovannetti for guidance and fruitful discussions. 
M.B. thanks the Scuola \mbox{Normale Superiore} in Pisa, Italy, for hosting him during a sabbatical that led to the development of this project. 
\end{acknowledgments}
\clearpage
%

\end{document}